\documentclass[conference,a4paper]{APSIPA2020}
\usepackage{multirow}
\usepackage{graphicx}
\usepackage{amsmath}
\usepackage{amssymb}
\usepackage{amsxtra}
\usepackage{threeparttable}
\usepackage{booktabs}
\usepackage{tabularx}
\usepackage[numbers,sort&compress,square]{natbib}
\usepackage{placeins}
\usepackage{lipsum}
\usepackage{cleveref}
\usepackage{siunitx}
\usepackage{subfigure}
\usepackage{ragged2e}
\usepackage{xcolor}
\usepackage{flushend} 
\usepackage[outdir=./figures/]{epstopdf}
\usepackage{jabbrv}
\crefname{figure}{Fig.}{Fig.}
\Crefname{figure}{Fig.}{Fig.}

\makeatletter
\def\blfootnote{\gdef\@thefnmark{}\@footnotetext}
\makeatother
\begin{document}

\title{A Strongly-Labelled Polyphonic Dataset of\\Urban Sounds with Spatiotemporal Context}

\author{%
\authorblockN{%
    Kenneth Ooi, 
    Karn N. Watcharasupat, 
    Santi Peksi, 
    Furi Andi Karnapi, 
    Zhen-Ting Ong, \\
    Danny Chua,
    Hui-Wen Leow,
    Li-Long Kwok, 
    Xin-Lei Ng, 
    Zhen-Ann Loh, 
    and Woon-Seng Gan
}
\authorblockA{
    School of Electrical and Electronic Engineering,
    Nanyang Technological University, Singapore\\
    Emails: \{wooi002, karn001, speksi, furi, ztong, jchua056, leow0121, e180115, xng033, zloh011, ewsgan\}@ntu.edu.sg}
}

\maketitle

\blfootnote
    {
    Part of this work were done while D. Chua, H.-W. Leow, L.-L. Kwok, X.-L. Ng, and Z.-A. Loh were with the School of Electrical and Electronic Engineering,
    Nanyang Technological University, Singapore.
}

\begin{abstract}
    This paper introduces {\textsc{\mbox{SINGA:PURA}}}, a strongly labelled polyphonic urban sound dataset with spatiotemporal context. The data were collected via several recording units deployed across Singapore as a part of a wireless acoustic sensor network. These recordings were made as part of a project to identify and mitigate noise sources in Singapore, but also possess a wider applicability to sound event detection, classification, and localization. This paper introduces an accompanying hierarchical label taxonomy, which has been designed to be compatible with other existing datasets for urban sound tagging while also able to capture sound events unique to the Singaporean context. This paper details the data collection, annotation, and processing methodologies for the creation of the dataset. We further perform exploratory data analysis and include the performance of a baseline model on the dataset as a benchmark.
\end{abstract}

\section{Introduction}
The collection of high-quality data for domain-specific problems is a major challenge in deep learning, and particularly so for tasks in the audio domain. Publicly available audio datasets captured in real-world contexts are few and far in between, owing to the high monetary and technical requirements needed to embark on audio data acquisition projects. As a result, synthetic data generation is often employed using software libraries such as Scaper \cite{Salamon2017Scaper:Augmentation} and Pyroomacoustics \cite{Scheibler2018Pyroomacoustics:Algorithms} in order to meet the training requirements of deep learning systems. Although these software libraries have been extremely helpful in the experimentation and development of many audio-based deep learning systems, synthetic data can never fully replicate the nuances of sounds recorded in real acoustic environments. The resulting covariate shift is fatal in many deep learning systems, thus leading to well-known problems of performance degradation when systems trained on synthetic data are deployed in the real world. Beyond generic sound event detection or classification system, urban sound tagging systems are amongst the applications in which real data are of extreme importance.

Audio data have unique distributions which are highly dependent not only on the sound sources themselves, but also on factors including the sonic environments and recording equipment. Moreover, when the sound itself is dependent on human activities, as is the case in urban sounds, further context (like the location and time of the recording) is often useful for both humans and machines to fully understand the sonic environment \cite{Cartwright2020SONYC-UST-V2:Context}. Unsurprisingly, the complex and layered nature of urban sounds means that synthetic samples are unable to perfectly replicate the real-world distribution of the frequency and timing of a given collection of urban sound sources. Hence, synthetic training samples are inadequate as training data for neural urban sound tagging systems. To develop systems which are able to cope with the complexity of the real world --- such as those in wireless acoustic networks deployed in Singapore \cite{Tan2021ExtractingSystem}, Germany \cite{Abeer2018StadtlarmEnvironment, Abeer2019UrbanReport}, Spain \cite{Salvo2020ASounds}, and the United States \cite{Cartwright2019SONYCNetwork} --- training with real-world data is vital.

At the time of writing, there exists no publicly available, strongly labelled urban sound dataset consisting of recordings in real-world contexts. This paper introduces the \mbox{SINGA:PURA} (\textit{Singapore: Polyphonic Urban Audio}) dataset\footnote{The dataset and detailed documentation are publicly available at https://doi.org/10.21979/N9/Y8UQ6F. In addition to the strongly labelled recordings, we have also included 72406 unlabelled recordings as part of the dataset.}, which is an urban sound dataset of 6547 strongly labelled recordings made in real acoustic environments with the corresponding spatiotemporal metadata attached to each recording. These recordings were made by single-channel sensors and multi-channel sensor arrays deployed across the island of Singapore via a wireless acoustic sensor network described in \cite{Tan2021ExtractingSystem}. In addition to urban sound tagging, this dataset would provide new grounds for applications of more complex tasks, such as sound event localization and detection, in urban settings.

\section{Existing Urban Sound Datasets}
Although there are a number of datasets for generic sound classification \cite{Gemmeke2017AudioEvents, Fonseca2017FreesoundDatasets, Piczak2015ESC:Classification, Mesaros2018AClassification, Trowitzsch2019} and synthetic datasets for urban sound classification \cite{Abeer2021USM-SEDScenarios, Johnson2021DESED-FLDetection}, there are currently only two main non-synthetic datasets of urban sound recordings that are publicly available. Specifically, they are the UrbanSound datasets \cite{Salamon2014AResearch}, and the SONYC Urban Sound Tagging datasets \cite{Bello2019SONYC:Pollution, Cartwright2020SONYC-UST-V2:Context}. We briefly discuss these datasets, as well as the related FSD50k dataset, in this section.

\subsection{Freesound FSD50k Dataset}
Although not an urban sound dataset by design, FSD50k \cite{Fonseca2017FreesoundDatasets} is to date one of the most comprehensive publicly available audio datasets. The dataset consists of over 100 hours of sounds drawn from Freesound, which is a database of audio clips licensed under various Creative Commons licenses. The recordings were annotated based on the AudioSet ontology \cite{Gemmeke2017AudioEvents}. Many smaller and task-specific datasets have been derived from Freesound, including the UrbanSound datasets \cite{Salamon2014AResearch}. 

However, since the audio data were crowdsourced and users on the Freesound platform are free to enter anything (or nothing) to describe their recordings, there is no proper documentation on the recording conditions, devices, and methods for these data. Moreover, the audio formats and number of channels vary across the entire dataset. Hence, these limitations apply to all datasets derived from FSD50k, such as the UrbanSound datasets, as well.

\subsection{UrbanSound and UrbanSound8k Datasets}
UrbanSound and UrbanSound8k are amongst the first datasets designed specifically for urban sound tagging \cite{Salamon2014AResearch}. They were curated using recordings from Freesound \cite{Fonseca2017FreesoundDatasets}. Both variants of UrbanSound consist of 10 coarse-level sound classes, namely air conditioner, car horn, children playing, dog bark, drilling, engine idling, gun shot, jackhammer, siren, and street music. The main UrbanSound dataset consists of 1302 potentially polyphonic recordings with only events from one class labelled strongly in each recording. UrbanSound8k contains 8732 labeled monophonic sound excerpts.

\subsection{SONYC Urban Sound Tagging Datasets}
The SONYC Urban Sound Tagging (SONYC-UST) dataset is a weakly labelled audio dataset of urban sound recordings. The audio was recorded by an urban acoustic sensor network under the Sounds of New York City (SONYC) project  \cite{Cartwright2019SONYCNetwork}. Early versions of the dataset only included the audio data \cite{Bello2019SONYC:Pollution}. Starting from version 2, the dataset also includes coarse spatiotemporal context (STC) of the location and time that each recording was made \cite{Cartwright2020SONYC-UST-V2:Context}.

The IEEE AASP Challenge on Detection and Classification of Acoustic Scenes and Events (DCASE) featured urban sound tagging as Task 5 in the 2019 and 2020 editions. The 2019 challenge utilized an early version (v0.2.0 and v0.3.0) of the dataset without spatiotemporal context. Recognizing the importance of spatiotemporal context in urban sound tagging, the 2020 challenge featured the second version (v2.2.0 and v2.3.0) of the dataset. Top-ranked systems from the 2020 challenge \cite{Arnault2020CRNNsContext, Iqbal2020IncorporatingTagging} demonstrated the efficacy of incorporating spatiotemporal features into deep learning systems for the classification of audio events whose distributions are context-dependent, such as that in urban sound tagging.

\section{The \mbox{SINGA:PURA} Dataset}

This section will describe the data acquisition, annotation, and processing methodologies for the creation of the \mbox{SINGA:PURA} dataset. 

\subsection{Data Acquisition}
\begin{figure}[tb]
    \centering
    \includegraphics[
        clip,
        height=4.5cm
    ]{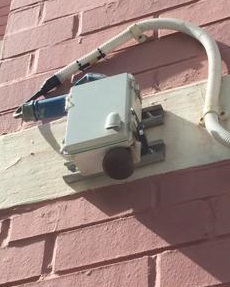}
    \includegraphics[
        clip,
        height=4.5cm
    ]{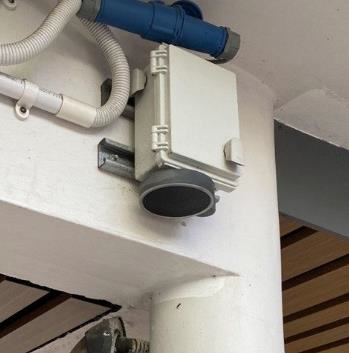}
    \caption{Sensor units used to make recordings for the SINGA:PURA dataset. \textit{Left}: A unit with a single-channel SPH0645 microphone. \textit{Right}: A unit with a seven-channel UMA-8 v2 microphone array.}
    \label{fig:dcu}
\end{figure}

Our data collection effort utilizes the wireless acoustic sensor network first introduced and detailed in \cite{Tan2021ExtractingSystem}. Each sensor unit is located in a public location, and houses either a single-channel Knowles SPH0645 microphone or a seven-channel MiniDSP UMA-8 v2 microphone array. The microphone in each sensor unit is controlled by a Raspberry Pi 3 Model B, which doubles a data collection unit for spatiotemporal information. All was recorded at a 44.1 kHz sampling rate. Pictures of the sensor units in their actual deployment contexts can be seen in \Cref{fig:dcu}. 

The recordings in the SINGA:PURA dataset are collected from 14 individual sensors deployed across four neighborhoods in Singapore, and span the duration between 3 August 2020 and 31 October 2020. The four neighborhoods consisted of two in the eastern region of Singapore and two in the western region of Singapore. The breakdown of the deployment by neighborhood, as well as the type of microphone is shown in \Cref{tab:deploy}.

\begin{table}[tb]
    \centering
    \caption{Data Collection Unit Deployment}
    \begin{tabular}{llr}
        \toprule
            \bfseries Neighborhood & 
            \bfseries Microphone & 
            \bfseries No. of Units \\
        \midrule
            East 1    & SPH0645  & 2 \\ 
            East 2    & SPH0645  & 3 \\ 
            West 1    & UMA-8 v2 & 4 \\ 
            West 2    & SPH0645  & 5 \\ 
        \bottomrule
    \end{tabular}
    \label{tab:deploy}
\end{table}

\subsection{Label Taxonomy}
The label taxonomy was derived from the taxonomy used in the SONYC-UST datasets \cite{Bello2019SONYC:Pollution, Cartwright2020SONYC-UST-V2:Context}, but has been adapted to fit the local context while retaining compatibility with the SONYC-UST ontology. Hence, this allows the SINGA:PURA dataset to be used in conjunction with the SONYC-UST datasets when training urban sound tagging models by simply omitting the labels that are absent in the SONYC-UST taxonomy from the recordings in the SINGA:PURA dataset.

\begin{figure*}[t!]
    \centering
    \includegraphics[width=0.9\textwidth]{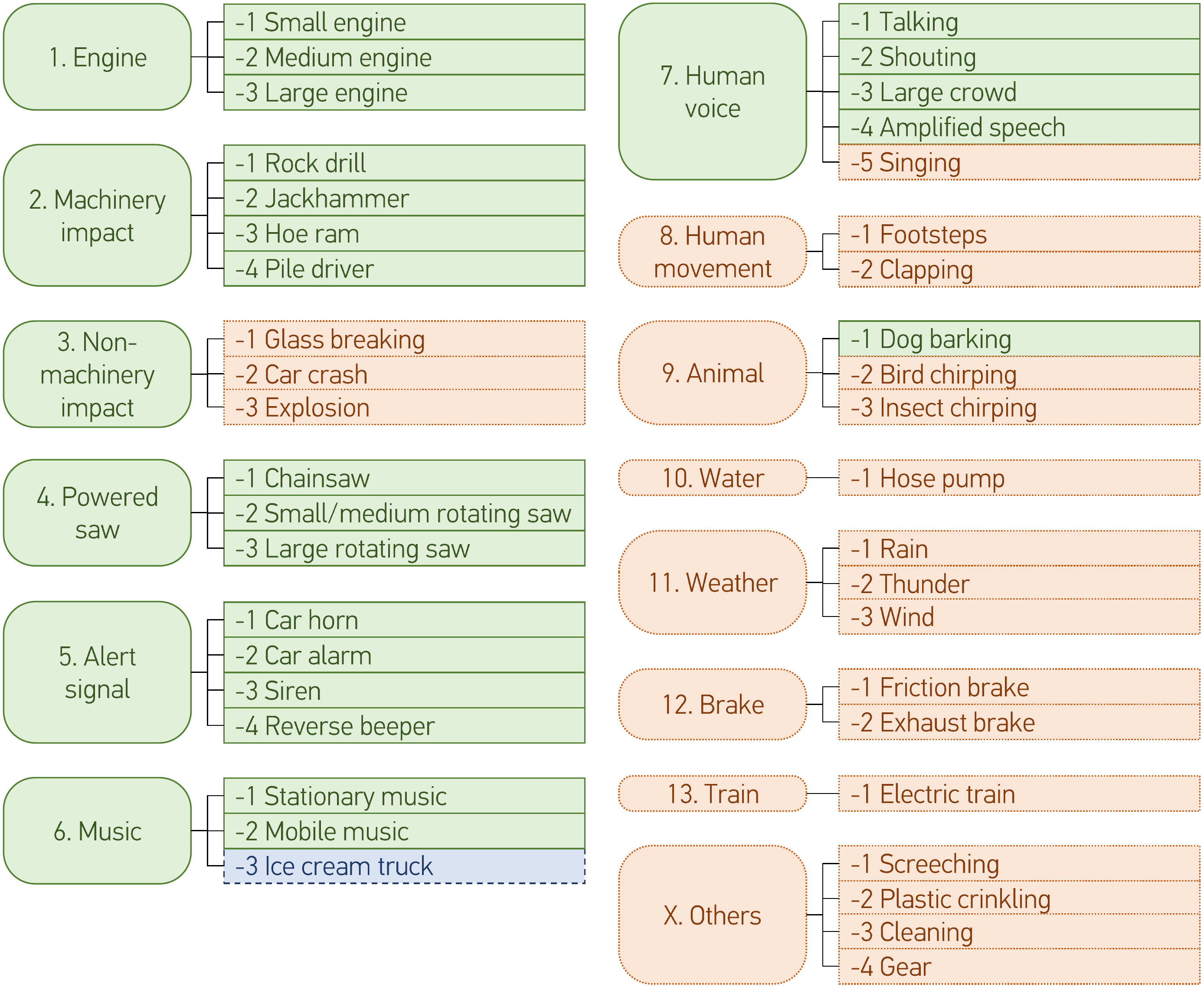}
    \caption{Label taxonomy for the \mbox{SINGA:PURA} dataset. Green boxes with solid outlines are classes common to both \mbox{SINGA:PURA} and SONYC-UST ontology. Red boxes with dotted outlines are classes newly introduced in \mbox{SINGA:PURA} ontology. The blue box with dashed outline indicates the class included in SONYC-UST ontology but excluded from \mbox{SINGA:PURA} ontology}
    \label{fig:taxonomy}
\end{figure*}

Our adapted taxonomy retains the original 8 coarse classes from SONYC-UST and introduces 6 new coarse classes, each possibly subsuming new fine classes as well. In addition, we have included a catch-all ``Others'' class to capture sound events that exist in the SINGA:PURA dataset, but do not fit well into any of the coarse class labels. Our taxonomy also preserves all the fine classes in the SONYC-UST ontonlogy except for the ``ice-cream truck'' class, since ice cream trucks do not exist in Singapore. In summary, our label taxonomy consists of 14 coarse classes and 40 fine classes, as illustrated in \Cref{fig:taxonomy}.

\subsection{Annotation}
The recorded audio data were split into 10-second chunks, before being individually annotated by a team of five annotators. Recordings with audible sensor faults were firstly excluded from the published dataset before being sent to the annotators to undergo the annotation process. The annotation process included identification of sound events, their fine class labels, their onset times, their offset times, and their estimated proximity from the sensor, thus resulting in polyphonic strong labels of all identifiable sound events in each recording. Each 10-second recording was annotated by at least one annotator, but some recordings were independently annotated by multiple annotators. In the case of recordings where the class labels of the sound events were ambiguous sound events, all annotators listened and a consensus decision was made to determine the sound classes, onsets, and offsets of the sound events.

All listening was done via Audio Technica ATH-M50x headphones, with playback levels matching that of white noise at an A-weighted equivalent sound pressure level of 65 dB. Calibration to the playback levels was performed on a GRAS 45BB KEMAR Head and Torso Simulator using the automatic calibration method described in \cite{Ooi2021AutomationHead}.

\subsection{Dataset structure}
The audio recordings are provided in the Free Lossless Audio Codec (FLAC) format at the original sampling rate of 44.1 kHz and are organized into folders according to the dates when the recordings were made.

The sound event labels are provided as a comma-separated values (CSV) file. The CSV file is organized in a similar manner to the label files used in DCASE sound event detection tasks, such as that used for DCASE 2021 Task 4 \cite{Turpault2019SoundSynthesis}. Each row in the label file represents exactly one sound event. The columns indicate the name of the recording, sound class, proximity, onset, and offset. A recording may contain multiple sound events and these are represented by multiple rows in the label file accordingly.

The spatiotemporal metadata is also provided as a separate CSV file, with each row containing the spatiotemporal context for each recording. Specifically, the columns in this CSV file indicate the sensor unit identifier, the date and time of recording, the day of the week, and the neighborhood the sensor was deployed in.

\Cref{tab:compare} shows a comparison between the \mbox{SINGA:PURA} dataset and other existing non-synthetic datasets.

\begin{table*}[htb]
    \centering
    \caption{Comparison of Non-Synthetic Urban Sound Datasets}
    \begin{tabular}{llrrcclr}
        \toprule
            &&\multicolumn{2}{c}{\bfseries No. of classes}\\
        \cmidrule{3-4}
            \bfseries Dataset &
            \bfseries Location & 
            \bfseries Coarse & 
            \bfseries Fine &
            \bfseries STC &
            \bfseries No. of Channels &
            \bfseries Label Type &
            \bfseries Duration (hours)\\
        \midrule
        \mbox{SINGA:PURA} &
            Singapore &
            14 & 40 &
            $\bullet$ &
            1 or 7 &
            Strong polyphonic &
            18.2 \\
        SONYC-UST v2.3 \cite{Cartwright2020SONYC-UST-V2:Context} &
            New York City &
            8 & 23 &
            $\bullet$ &
            1 &
            Weak polyphonic &
            51.4 \\
        SONYC-UST v1.0 \cite{Bello2019SONYC:Pollution} &
            New York City &
            8 & 23 &
            $\times$ &
            1 &
            Weak polyphonic &
            8.5 \\    
        UrbanSound \cite{Salamon2014AResearch} &
            Varies, unknown &
            10 & $\times$ &
            $\times$ &
            varies &
            Strong monophonic &
            18.5\\
        UrbanSound8k \cite{Salamon2014AResearch} &
            Varies, unknown &
            10 & $\times$ &
            $\times$ &
            varies &
            Weak monophonic &
            9.7\\
        \bottomrule
    \end{tabular}
    \label{tab:compare}
\end{table*}

\section{Exploratory Data Analysis}

In total, 6547 recordings of 10 seconds were annotated, totaling to 18.2 hours of audio data. Of these, 38 recordings were found to have no sound events in the taxonomy illustrated in \Cref{fig:taxonomy}. All fine-level class labels have at least one example in the other 6509 recordings in the dataset,  \textit{except} for the ``car crash'' class (class 3-2). Hence, only 39 of the 40 fine-level sound classes in \Cref{fig:taxonomy} will be represented in the current version of the \mbox{SINGA:PURA} dataset.

\begin{figure}[tb]
    \centering
    \includegraphics[width=0.73\columnwidth]{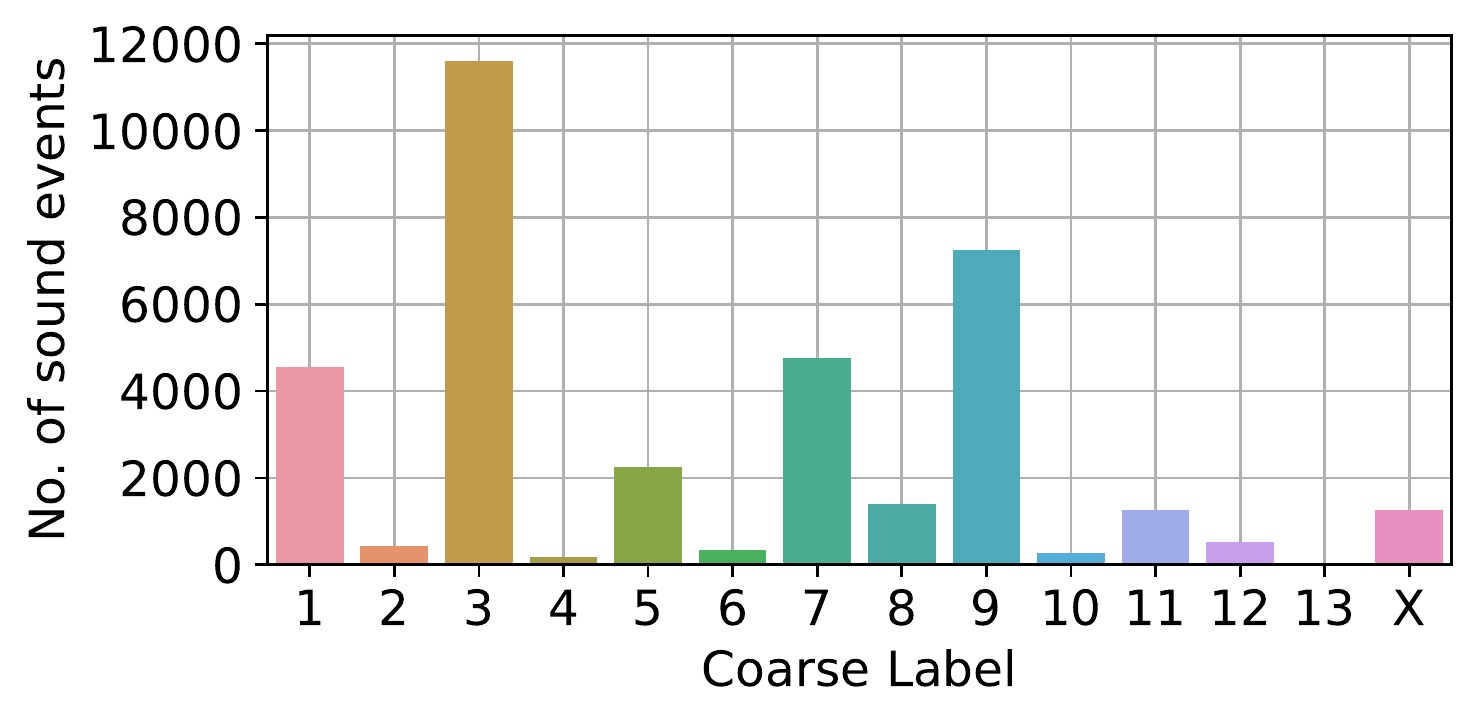}
    \vspace{-1em}
    \caption{Distribution of the number of disjoint sound events by coarse-level class}
    \label{fig:count}
    \includegraphics[width=0.73\columnwidth]{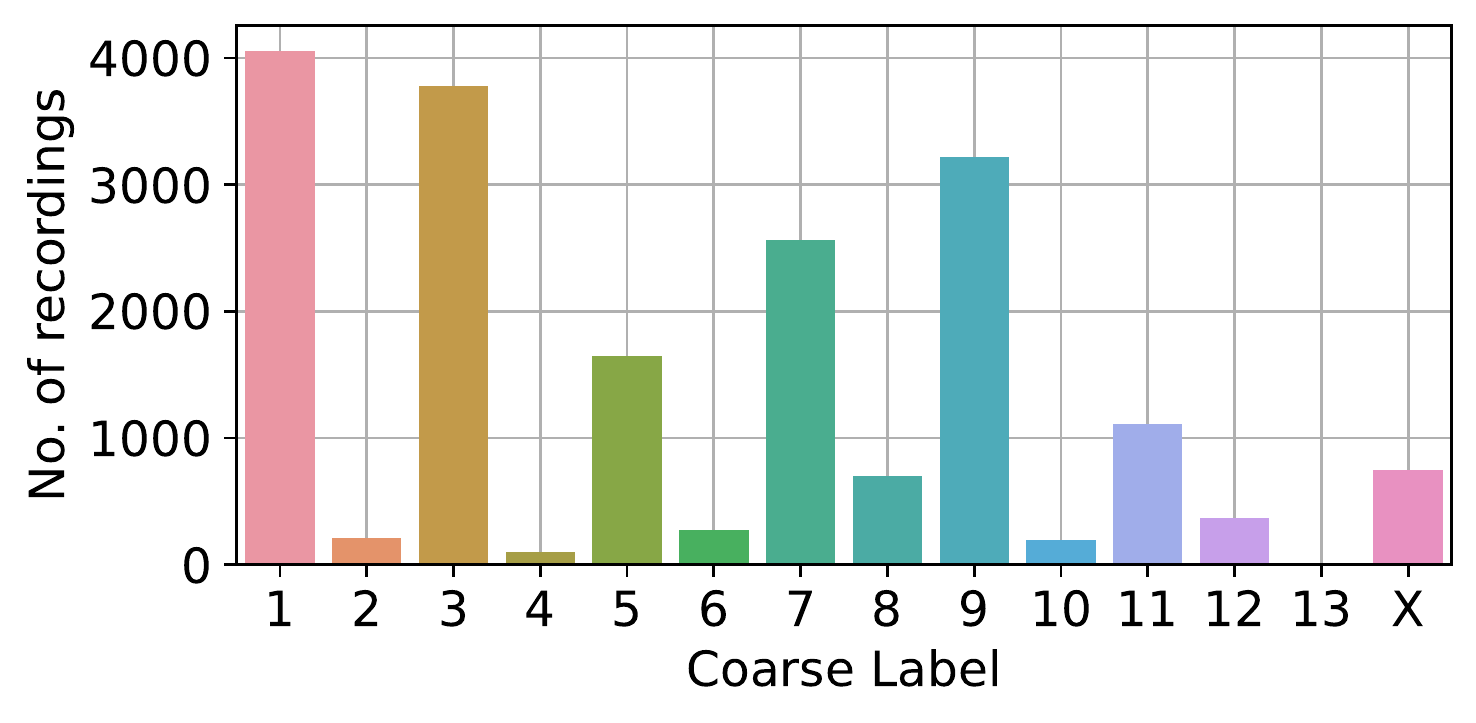}
    \vspace{-1em}
    \caption{Distribution of the number of recordings with at least one sound event by coarse-level class}
    \label{fig:rcount}
    \includegraphics[width=0.73\columnwidth]{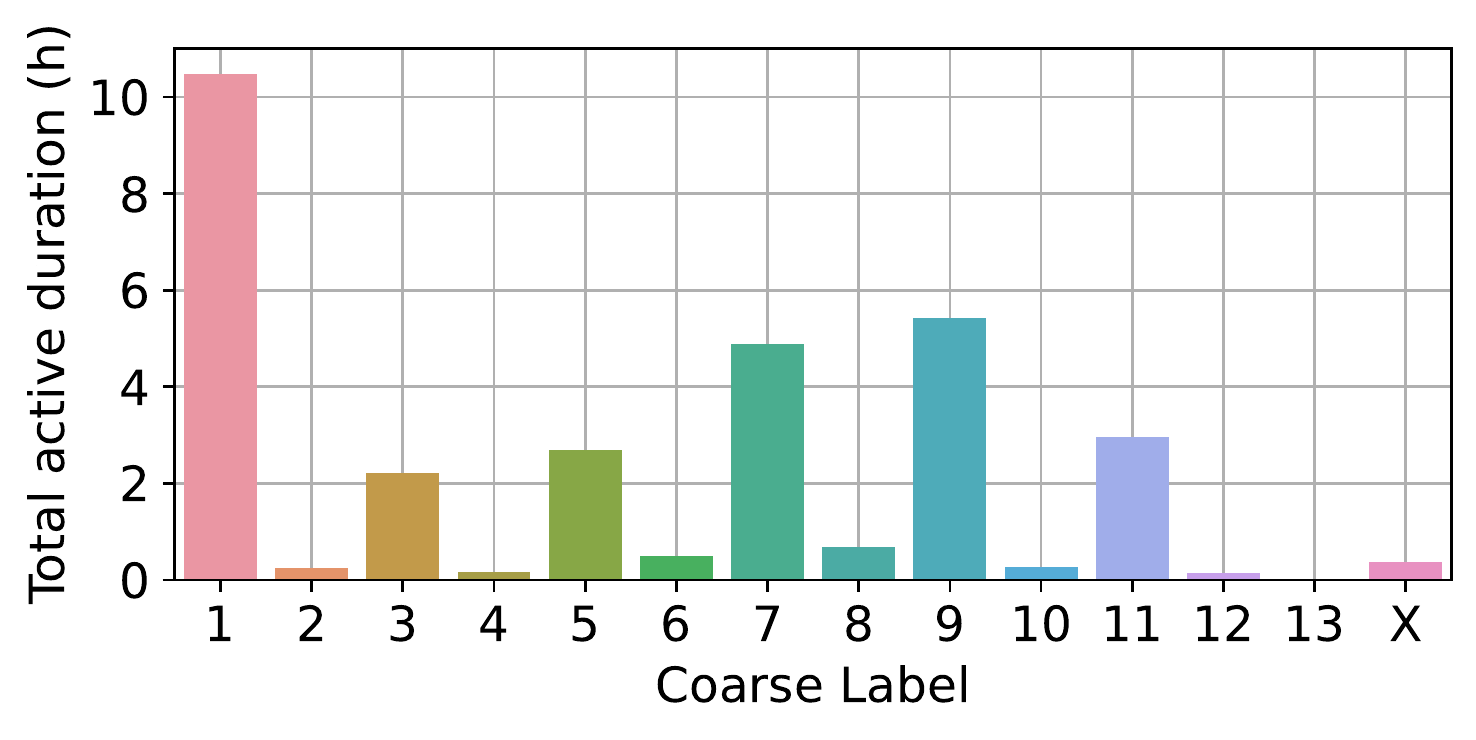}
    \vspace{-1em}
    \caption{Distribution of the total duration of active sound events by class}
    \label{fig:dur}
    \includegraphics[width=0.73\columnwidth]{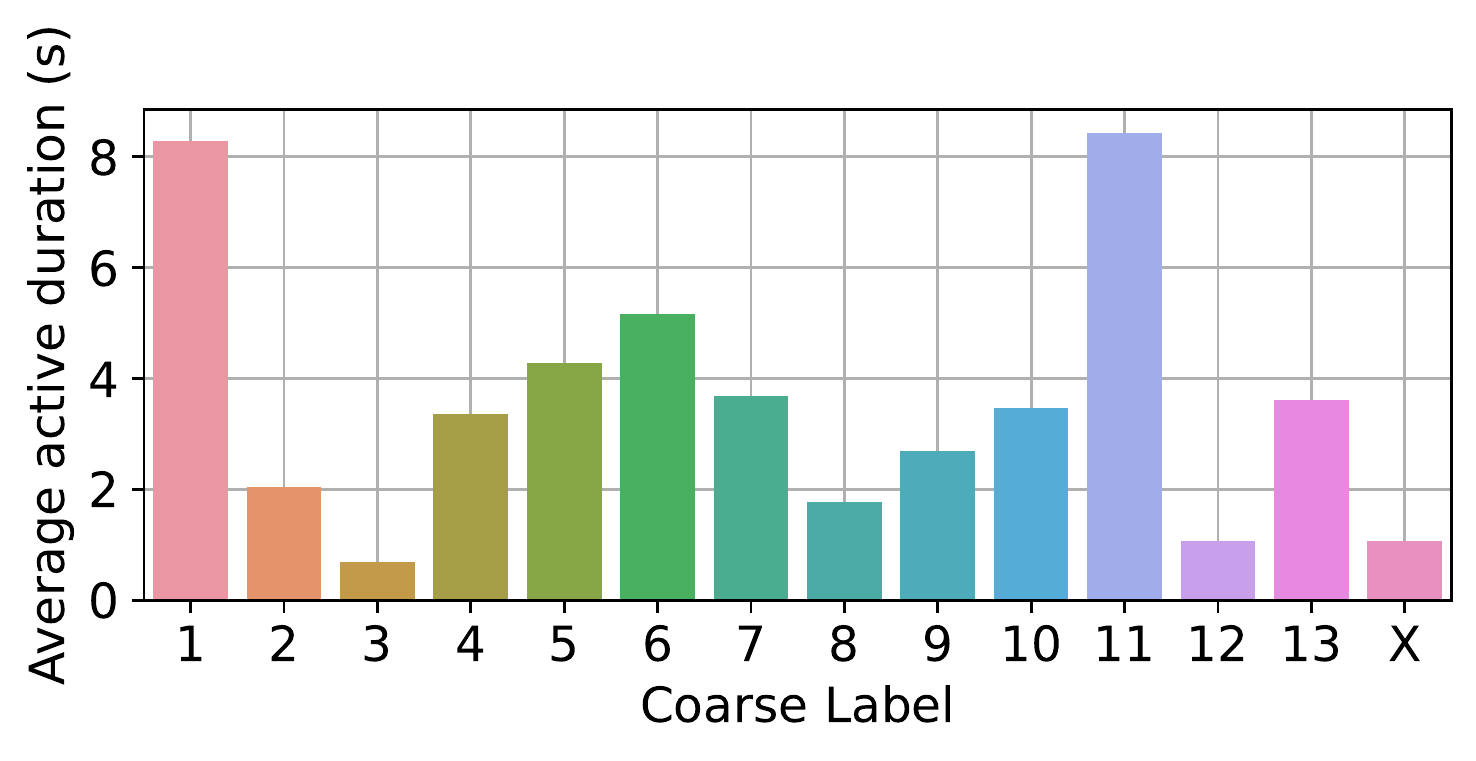}
    \vspace{-1em}
    \caption{Average sound event duration by coarse-level class}
    \label{fig:avg}
    \includegraphics[width=0.73\columnwidth]{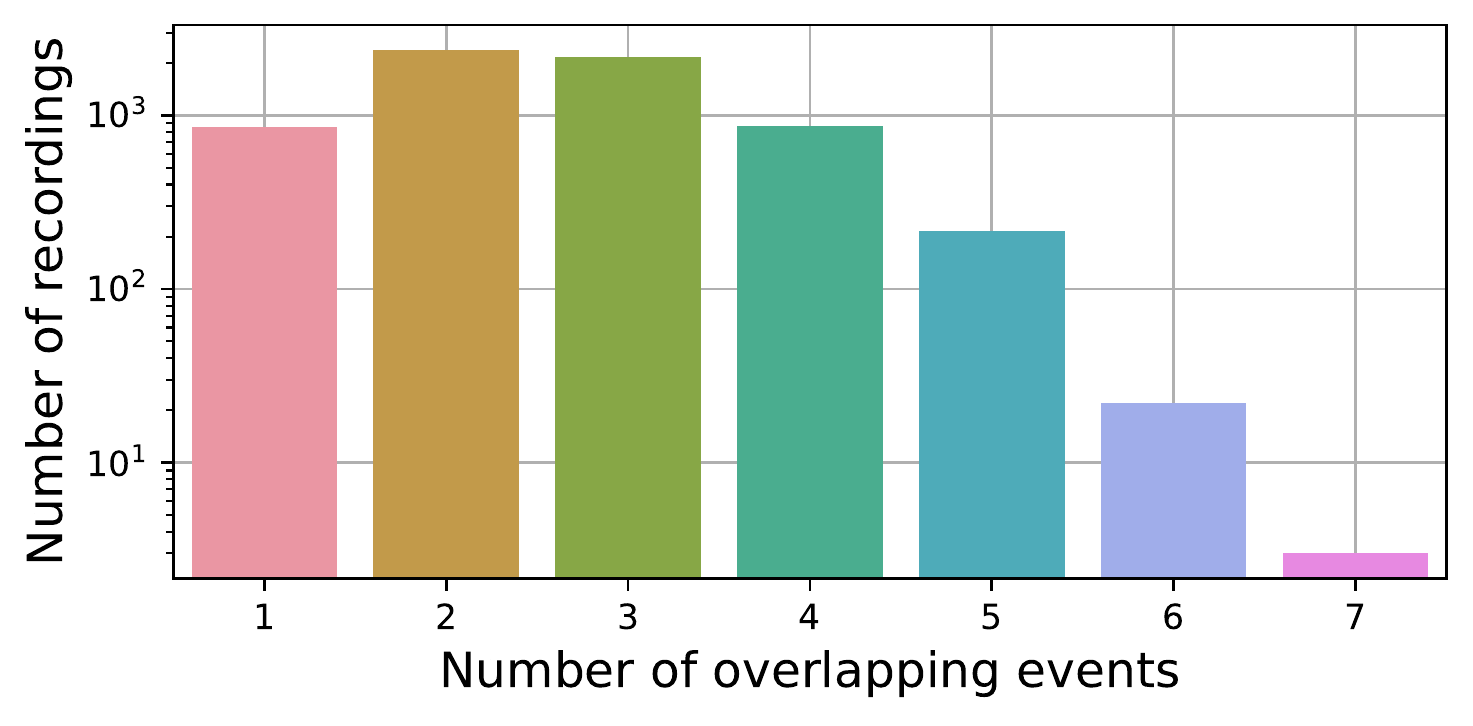}
    \vspace{-1em}
    \caption{Distribution of the maximum number of overlapping sound events at any point in time in a recording.}
    \vspace{-1em}
    \label{fig:poly}
\end{figure}

The distribution of the individual sound events by class is shown in \cref{fig:count}. Sound events are considered as continuous intervals of time, so it is possible for a single recording to contain two sound events of the same class, even if the time intervals when they are active are overlapping. We can see that the ``non-machinery impact'' class (class 3) has the highest frequency of occurrence, followed by the ``animal'' class (class 9). The high occurrence of the ``animal'' class is largely due to the presence of cicadas and crickets chirping in a large number of recordings, since they are very commonly found in Singapore.

In contrast, the distribution of the number of recordings with at least one sound event by coarse-level class is shown in \cref{fig:rcount}. We can see that certain sound classes, such as the ``engine'' class (class 1) and the ``non-machinery impact'' class (class 3), appear in many recordings. Since a given recording may have multiple sound events, the total number shown in all the bars exceeds the total number of recordings (6547) in the SINGA:PURA dataset.

The distribution of the total active duration of sound events by coarse-level class across the entire dataset is shown in \cref{fig:dur}. We can see that the distribution is unbalanced across the coarse classes and follows an inverse power law, which is typical of real recordings in urban environments. In a similar manner as \cref{fig:rcount}, the total duration represented by all the bars in \cref{fig:dur} is longer than that of all the recordings in the SINGA:PURA dataset because each recording may have multiple sound events that overlap.

\Cref{fig:avg} shows the average active duration of sound events by coarse-level class. As expected, sound classes that are characterized by stationary or ambient sounds, such as the ``engine'' class (due to the presence of traffic) and the ``weather'' class (due to the presence of rain), naturally have longer average durations compared to other classes. On the other hand, sound classes characterized by impulsive sounds, such as the ``non-machinery impact'' class (class 3) and the ``brake'' class (class 12) naturally have shorter average durations compared to other classes.

\Cref{fig:poly} shows the distribution of the maximum number of overlapping sound events at any point in time in a recording. The most common degree of polyphony in the dataset is two, while the maximum degree of polyphony is seven. Beyond a degree of polyphony of three, the distribution shows an exponential decrease, as can be seen form the logarithmic scale used in the y-axis of \Cref{fig:poly}.

\begin{figure}[tb]
    \centering
    \includegraphics[width=\columnwidth]{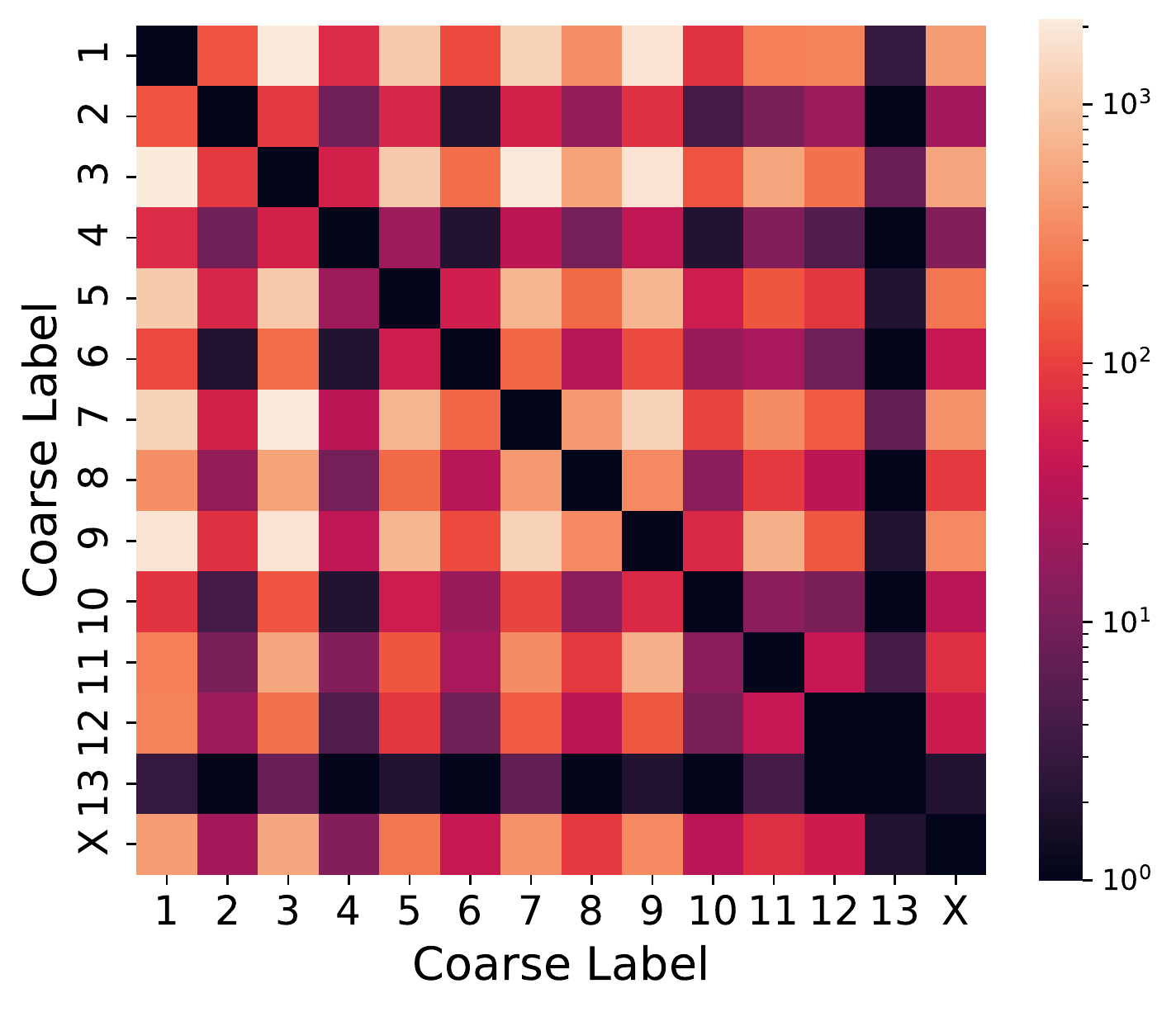}
    \caption{Coarse-level co-occurrence matrix computed at recording level. Note that the color bar is in the logarithmic scale.}
    \label{fig:coocc}
\end{figure}

Lastly, \Cref{fig:coocc} shows the co-occurrence matrix of the coarse-level sound classes at the recording level. It can be seen that certain classes of sound events tend to occur together in a recording (e.g. engine, non-machinery impact, alert signal, human voice, and animal), while some classes of sound events rarely or never occur together in the same recording (e.g. water with either machinery impact or powered saw). 

\section{Baseline Model}
As a baseline model, we adapted the baseline model from Task 5 (Urban Sound Tagging with Spatiotemporal Context) of the DCASE 2020 Challenge \cite{Cramer2020BaselineChallenge} for the \mbox{SINGA:PURA} dataset. The adapted system and training process are detailed in \cite{Kwok2021UrbanContext} and we replicate the results here for ease of reference. 
The system in \cite{Kwok2021UrbanContext} is identical to the model in \cite{Cramer2020BaselineChallenge} except for the prediction layer which was changed to accommodate the additional sound classes in the \mbox{SINGA:PURA} dataset. The model was initially trained on the SONYC-UST v2.2 dataset \cite{Cartwright2020SONYC-UST-V2:Context} with coarse labels. The original output layer was then detached and replaced with a 14-class output layer. The model was then fine-tuned for coarse-level prediction on the \mbox{SINGA:PURA} dataset. 
The overall and class-wise performances of the model are shown in \Cref{tab:results}. For the summary of results, we report the micro- and marco-aggregated areas under the precision-recall curve (AUPRC) as well as a micro-aggregated F1 score. For the class-wise results, we report the AUPRC. Although the results cannot (and should not) be directly compared in a conventional sense, we also include the performance of the coarse-level model trained on the SONYC-UST v2.2 dataset from \cite{Cramer2020BaselineChallenge} as a rough reference.

\begin{table}[tb]
    \centering
    \caption{Baseline Model Performance}
    \begin{tabular}{lSS}
        \toprule
            \bfseries Metrics & 
            \bfseries \mbox{SINGA:PURA} &
            \bfseries SONYC\textendash UST\\
        \midrule
            Micro AUPRC &   0.7064  &  0.8352\\
            Micro F1    &   0.4811  &  0.6736\\
            Macro AUPRC &   0.5825  &  0.6370\\
        \midrule
            Class-wise AUPRC\\
            \quad 1. Engine                 &   0.8706  & 0.8500\\
            \quad 2. Machinery impact       &   0.2284  & 0.6021\\
            \quad 3. Non-machinery impact   &   0.8553  & 0.4192\\
            \quad 4. Powered saw            &   0.0220  & 0.7200\\
            \quad 5. Alert signal           &   0.7679  & 0.8518\\
            \quad 6. Music                  &   0.4890  & 0.6145\\
            \quad 7. Human voice            &   0.8708  & 0.9593\\
            \quad 8. Human movement         &   0.5354\\
            \quad 9. Animal/Dog*            &   0.7906  & 0.0463\\
            \quad 10. Water                 &   0.1682\\
            \quad 11. Weather               &   0.7698\\
            \quad 12. Brake                 &   0.1649\\
            \quad 13. Train                 &   0.0000\\
            \quad X. Others                 &   0.2029\\
        \bottomrule
    \end{tabular}
    \label{tab:results}
    \begin{justify}
    * The SONYC-UST taxonomy does not have an \textit{animal} coarse-level class but has a coarse-level \textit{dog} class. In the SINGA:PURA taxonomy, the \textit{dog} class is a fine-level class under the \textit{animal} class. We include the result from the \textit{animal} class for SINGA:PURA and from the \textit{dog} class for SONYC-UST.
    \end{justify}
\end{table}

\section{Conclusions}
This paper introduces a non-synthetic polyphonic urban sound dataset with spatiotemporal context and strong labels. The data collection was done via a wireless acoustic sensor network deployed across Singapore. The multi-channel nature of a number of the recordings from the dataset also opens the possibility of localizing sound events within the dataset on top of detecting or classifying them. Lastly, an expanded label taxonomy for urban sounds compatible with the SONYC-UST taxonomy was introduced, thus allowing for benchmarking of urban sound tagging systems across multiple cities. 

\FloatBarrier
\section*{Acknowledgment}
This research is supported by the National Research Foundation and the Smart Nation Digital Government Office, Prime Minister’s Office, Singapore, under the Translational Research and Development for Application to Smart Nation (TRANS) Grant. Any opinions, findings, and conclusions or recommendations expressed in this material are those of the authors and do not reflect the views of the National Research Foundation and the Smart Nation Digital Government Office, Prime Minister’s Office, Singapore.


K. N. Watcharasupat acknowledges the support from the CN Yang Scholars Programme, Nanyang Technological University, Singapore.

The authors thank Alroy Chiang from the Division of Physics and Applied Physics, School of Physical and Mathematical Sciences, Nanyang Technological University, Singapore, for his assistance with data processing.

\bibliographystyle{ieeetran}
\bibliography{refs.bib}

\end{document}